\documentstyle[preprint,aps,floats,psfig]{revtex}
\begin{document}
\def\mh{m_h^{}}
\def\vev#1{{\langle#1\rangle}}
\def\gev{{\rm GeV}}
\def\tev{{\rm TeV}}
\def\fbi{\rm fb^{-1}}
\def\lsim{\mathrel{\raise.3ex\hbox{$<$\kern-.75em\lower1ex\hbox{$\sim$}}}}
\def\gsim{\mathrel{\raise.3ex\hbox{$>$\kern-.75em\lower1ex\hbox{$\sim$}}}}
\newcommand{\hmu}{{\hat\mu}}
\newcommand{\hnu}{{\hat\nu}}
\newcommand{\hrho}{{\hat\rho}}
\newcommand{\hh}{{\hat{h}}}
\newcommand{\hg}{{\hat{g}}}
\newcommand{\hk}{{\hat\kappa}}
\newcommand{\tA}{{\widetilde{A}}}
\newcommand{\tP}{{\widetilde{P}}}
\newcommand{\tF}{{\widetilde{F}}}
\newcommand{\th}{{\widetilde{h}}}
\newcommand{\tp}{{\widetilde\phi}}
\newcommand{\tchi}{{\widetilde\chi}}
\newcommand{\te}{{\widetilde\eta}}
\newcommand{\vn}{{\vec{n}}}
\newcommand{\vm}{{\vec{m}}}

\newcommand{ \slashchar }[1]{\setbox0=\hbox{$#1$}   
   \dimen0=\wd0                                     
   \setbox1=\hbox{/} \dimen1=\wd1                   
   \ifdim\dimen0>\dimen1                            
      \rlap{\hbox to \dimen0{\hfil/\hfil}}          
      #1                                            
   \else                                            
      \rlap{\hbox to \dimen1{\hfil$#1$\hfil}}       
      /                                             
   \fi}                                             %

\tighten
\preprint{ \vbox{
\hbox{MADPH--00-1156}
\hbox{hep-ph/0001320}}}
\draft
\title{Oblique Parameter Constraints on Large Extra Dimensions}
\author{Tao Han, Danny Marfatia, and Ren-Jie Zhang}
\address{Department of Physics, University of Wisconsin\\ 
1150 University Avenue, Madison, WI 53706, USA}
\date{January 2000}

\maketitle

\begin{abstract}
{\rm  We consider the Kaluza-Klein scenario in which gravity 
propagates in the $4+n$ dimensional
bulk of spacetime and the Standard
Model particles are confined to a 3-brane. We calculate the gauge
boson self-energy corrections arising from the exchange of 
virtual gravitons and present our results in the $STU$-formalism. 
We find that the new physics contributions to $S$, $T$ and $U$ decouple
in the limit that the string scale $M_S$ goes to infinity. 
The oblique parameters constrain the lower 
limit on $M_S$. Taking the quantum gravity cutoff to be $M_S$, 
 $S$-parameter constraints impose $M_S>1.55$ TeV for 
$n=2$ at the 1$\sigma$ level. $T$-parameter constraints
impose \mbox{$M_S>1.25\ (0.75)$ TeV} for $n=3\ (6)$. 

}
\end{abstract}

\section{Introduction}
The possible existence of extra dimensions has been a fascinating
idea in physics ever since Kaluza-Klein theory was 
proposed \cite{kk}.
Consistent string theories demand the existence of extra 
dimensions \cite{strings}. However, if the string scale ($M_S$) 
is as high as the grand unification scale ($M_{\rm GUT}\sim 10^{16}$ GeV) 
or the Planck scale ($M_{\rm Pl}\sim 10^{19}$ GeV), as is the 
case for a weakly coupled heterotic string, then the length 
scale of the compactified extra dimensions $(R\sim 1/M_S)$ 
would be too small to be appreciable experimentally. 
Recent developments in string theory indicate that the string scale
can be much lower than the Planck scale and even close to the
electroweak scale \cite{ewstring}. 
This possibility provides new avenues towards many theoretical
issues such as alternative solutions to the gauge hierarchy
problem \cite{add,rs}, fermion mass and flavor mixings \cite{flavor},
and new inflationary cosmological models \cite{cosmo}. More importantly,
such a scenario may lead to a rich phenomenology and is thus
experimentally testable at low energies \cite{orgph,grw,hlz,pheno,virtual}.

Assume that there are $n$ extra dimensions in which only gravity
can propagate while Standard Model (SM) fields are confined to
four dimensional spacetime. The large value of the Planck scale 
can be understood by Gauss' law from the relation 
\begin{equation}
M_{\rm Pl}^2 \sim R^{n} M_S^{n+2} \ ,
\end{equation}
where the string scale $M_S$ is taken to be the Planck mass in $4+n$ 
dimensions. For \mbox{$M_S\sim{\cal O}$ (1 TeV)}, $R$ can range from 
1 fm to 1 mm for $n=6$ to 2 \cite{add}. There are no direct
gravitational tests sensitive to those small scales yet \cite{subm}. 
Such large extra dimensions manifest 
themselves only through interactions involving the 
Kaluza-Klein (KK) modes of the gravitons with enhanced coupling
strength after summing over the many contributing light KK
states. The effective theory governing the graviton 
couplings to matter was described in \cite{grw,hlz}. 
Phenomenological studies showed that future collider experiments
can provide constraints on $M_S$ typically of order \mbox{1 TeV},
 depending on the collider 
center of mass energies \cite{grw,pheno,virtual}. 
Astrophysical (cosmological) 
considerations have been used to impose lower bounds 
on $M_S$ to be about 30 (100) TeV for $n=2$ and
very weak bounds for higher dimensions \cite{bounds}.

In this paper we consider radiative corrections to the masses
of the electroweak gauge bosons $(W,Z)$ arising from the 
exchange of massive spin-2 KK gravitons. The motivation for
this study is two-fold. Firstly, a rigorously renormalizable
theory of gravity does not exist. It would be interesting 
to explore to what extent the formalism 
in Ref.~\cite{grw,hlz} is finite against radiative corrections
in the sense of an effective field theory. An early attempt
in Ref.~\cite{hlz} for a scalar self-energy correction
showed that the radiative corrections are proportional
to the scalar mass, instead of the ultraviolet cutoff.
Another one-loop calculation for the muon $g-2$ also reached 
finite results \cite{loopbds}. 
Secondly, if one is able to compute radiative
effects from  KK gravitons, one may hope to examine
 constraints on new physics characterized by the string 
scale from precision electroweak measurements. 
In fact, a recent paper appeared to estimate the 
 $\rho$ parameter \cite{wrong}, which was
found to be both ultraviolet and infrared divergent. However,
our results arrive at completely different conclusions.

The rest of the paper is organized as follows. In Section 
II we compute one-loop self-energy diagrams for a gauge boson
from exchange of massive spin-2 KK gravitons. To do so,
we need to derive the four--point couplings of two gravitons
and two gauge bosons which are beyond that given in 
Ref.~\cite{grw,hlz} and have been left out
in the previous calculations \cite{loopbds,wrong}.
In Section III we adopt the $STU$-formalism \cite{para}
and constrain the string scale based on
    current experimental values of the oblique parameters.
Computing oblique parameters has an advantage over 
calculating $W(Z)$ mass corrections directly since they
are free of certain technical and conceptual uncertainties.  
We discuss our results and conclude in Section IV. 
In two appendices we present the relevant Feynman rules 
and describe the regularization of the infrared divergences.

\section{Self-energy corrections to the gauge bosons}

\begin{figure}[b]
\centerline{\psfig{file=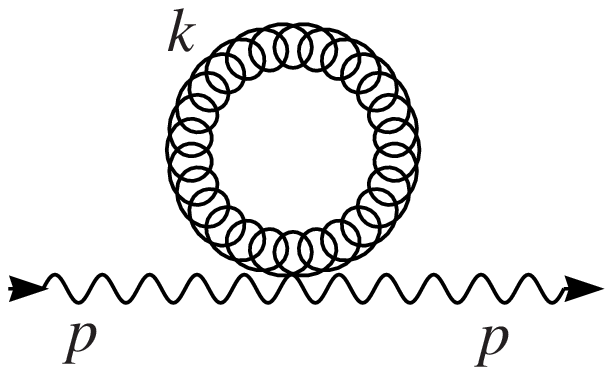,width=7cm,height=4cm}
\psfig{file=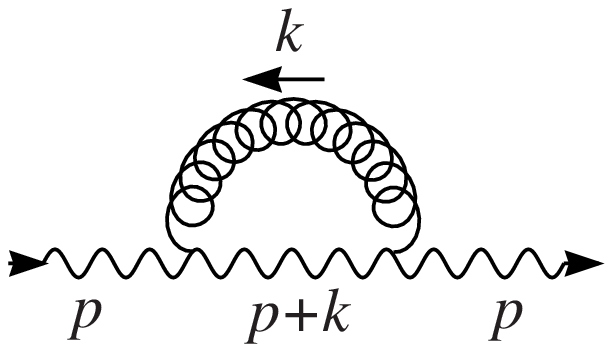,width=7cm,height=4cm}}
\bigskip
\caption[]{The seagull and rainbow diagrams contributing 
to the self-energy of gauge bosons. }
\label{graphs}
\end{figure} 

When the typical energy scale of interest is much smaller than
the string scale $M_S$, the compactified
higher dimensional theory can be described by an effective theory
where only relevant light degrees of freedom are retained, namely,
those related to the so called Kaluza-Klein states. In principle, if
only gravity is of higher dimensional origin, spin-0, 1
and 2 KK states can arise \cite{grw,hlz}.
The spin-1 states decouple from the Standard Model fields and make 
no contribution to the self-energies. Naively, one expects the spin-0
contributions to be of the same order of magnitude as that of the spin-2
states since they have the same coupling strength. However, the properties
of the spin-0 states are model-dependent since there is no symmetry to
protect their masses.
In the following we shall concentrate 
on the spin-2 KK states (which we will call KK gravitons).

The physical process that we consider in this paper is the 
radiative corrections to the weak gauge boson masses at the one-loop
level from these KK gravitons. 
We focus our attention on the transverse part of the self-energy
$\Pi_{XY}(p^2)$ between gauge bosons $X$ and $Y$, as it is
the only part relevant to us. These self-energies are written as
\begin{equation}
\Pi_{Z}(p^2)\equiv \Pi^T_{ZZ}(p^2)\quad {\rm and}\quad 
\Pi_{W}(p^2)\equiv \Pi^T_{WW}(p^2).
\end{equation}
Note that both $\Pi_{Z\gamma}$ and $\Pi_{\gamma\gamma}$ are identically zero,
following from the simple fact that they must be proportional to the 
photon mass, as required by the gravitational nature of the KK
graviton-matter interactions.

There are two diagrams contributing to the self-energy of a gauge boson,
as shown in Fig.~\ref{graphs}. The first (seagull diagram)
involves a four--point
coupling of two KK gravitons and two gauge bosons,
and the second (rainbow diagram) involves two three--point 
couplings of a KK graviton and two gauge bosons. 
These two diagrams are at the same order in
the gravitational coupling $\kappa^2=16\pi G_N$, where $G_N$ 
is the usual four-dimensional Newton's constant. 

To obtain the four-point vertex Feynman rule required for the seagull diagram,
one needs to expand the graviton-matter interaction
Lagrangian to order $\kappa^2$, which is beyond the order of the expansion 
given in \cite{hlz}. We provide this Feynman rule in Appendix A.
The corresponding seagull diagram is purely transverse, proportional
to $(\eta_{\mu\nu} - p_\mu p_\nu/p^2)\ \Pi_S(p^2)$, with
\begin{equation}
\Pi_{S}(p^2)\ =\ 
{\frac{\kappa^2p^2}{16 \pi^2}}\sum_{\vec{n}} \int_0^{\infty} 
{\frac{dk_E^2\, k_E^2}
{k_E^2+m_{\vec{n}}^2}} \left(
{\frac{k_E^4}{12m_{\vec{n}}^4}}+
{\frac{3k_E^2}{4m_{\vec{n}}^2}}+{3\over2}\right)\ ,
\end{equation}
where we have included a factor $1/2$ to avoid double-counting when
we sum over the KK modes  and
the subscript $E$ indicates that the variable 
is in Euclidean space. The summation 
over the KK modes in a tower can be written as an integration 
because of the near-degeneracy of the KK states,
\begin{equation}
\sum_{\vec{n}} f(m_{\vec{n}})= \int_0^{\infty} dm_{\vec{n}}^2 
\rho(m_{\vec{n}}) f(m_{\vec{n}})
\label{summation}
\end{equation}
where
\begin{equation}
\rho(m_{\vec{n}})={\frac{R^n m_{\vec{n}}^{n-2}}{(4 \pi)^{n/2}\, 
\Gamma(n/2)}}
\label{density}
\end{equation}
is the KK state density. By convention for the torus 
compactification \cite{hlz}, the relation between the 
four-dimensional Newton's constant and the $(4+n)$-dimensional 
string scale $M_S$ is given by
\begin{equation}
\kappa^2 R^n=8 \pi (4 \pi)^{n/2}\, \Gamma(n/2) M_S^{-(n+2)} .
\label{convention}
\end{equation}
Using Eqs.~(\ref{summation}),~(\ref{density}) 
and ~(\ref{convention}), we obtain
\begin{equation}
{\frac{\kappa^2}{16 \pi^2}} \sum_{\vec{n}} f(m_{\vec{n}})=
{\frac{1}{2\pi M_S^{n+2}}}
\int_0^{\infty} dm_{\vec{n}}^2\, 
m_{\vec{n}}^{n-2} f(m_{\vec{n}}^2).
\end{equation}

The above sum is divergent in the ultraviolet for $n\geq2$ 
unless $f(m_{\vec{n}})$ falls off very rapidly with $m_{\vec{n}}^2$.
Since the effective theory
is only expected to be valid below the string scale, we 
introduce an explicit cutoff $\lambda M_S$ to regularize the
mass sum, where $\lambda\sim {\cal O}(1)$ and parameterizes 
the sensitivity to the cutoff.\footnote{This bad ultraviolet 
behavior can be remedied in models of fluctuating branes
and models of fermions located in different branes \cite{ultra}.} 
%
We shall use the same cutoff for the momentum integral
as for the mass summation. We believe that
this regulator will best reflect the ultraviolet behavior
of the divergence. Our final result for the seagull diagram is
\begin{equation}
\Pi_{S}(p^2)={\frac{\lambda^{n+2}p^2}{12\pi}} 
\int_0^1\int_0^1 dxdy ~ {\frac{x\,{y^{-3 + 
{\frac{n}{2}}}}}{x+y}}\,
 \left( x + 3\,y \right) \,\left( x + 6\,y \right)\ ,
\label{seagull}
\end{equation}
where $x=(k_E/\lambda M_S)^2$ and  $y=(m_{\vec{n}}/\lambda M_S)^2$. 

The necessary Feynman rules for calculating the rainbow 
diagram have been derived in \cite{hlz} and are summarized 
in Appendix A. The complete expression
for the general $\Pi_R(p^2)$ 
is lengthy and unilluminating. We will instead present the
special cases relevant to our current consideration, $\Pi_R(0)$ and  
$\Pi_R(m^2)$. The former can be written as
%
\begin{eqnarray}
\Pi_R(0)&=& {\frac{\lambda^{n+2}m^2}{24 \pi}} \int_0^1\int_0^1
dxdy~ {\frac{x\,{y^{-3 + {\frac{n}{2}}}}}
   {\left( r + x \right)\left( x + y \right) }}\times\nonumber \\ 
&& \biggl[x\left( x + y \right) 
\left( x + 13y \right)  + 
  r\left( 4{x^2} + 26xy + 52{y^2} \right) \biggr]\ ,
\label{rainbow0}
\end{eqnarray}
where we have introduced a dimensionless mass ratio between
the gauge boson mass and the ultraviolet cutoff
$r=(m/\lambda M_S)^2$. Similarly,
%
%
\begin{eqnarray}
\Pi_R(m^2)&=& {\frac{\lambda^{n+2}m^2}{24 \pi}}  
\int_0^1\int_0^1\int_0^1 dx dy dz~
\,{\frac{x\,{y^{-3 + {\frac{n}{2}}}}\ F(x,y,z,r)}
   {{{[ x + y ( 1 - z )  + r\,{z^2} ]}^2}}}\ ,
\label{rainbowm}
\end{eqnarray}
where $z$ is a Feynman parameter and
\begin{eqnarray}
F(x,y,z,r)&=&  4\,{r^3}\,{{\left( -2 + z \right) }^2}\,{z^4}
       + 
  {r^2}\,{z^2}\,\biggl[ 16y ( -2 + z)  + 
     x ( -24 + 52z - 21{z^2})  \biggr]
\nonumber \\ & &
 + r\biggl[ 4xy ( 2 - 8z + {z^2} )  + 
     4{y^2} ( 4 + 9{z^2} )  + 
     {x^2} ( 1 - 14\,z + 15{z^2} )  \biggr] 
\nonumber \\ & &
 - x ( {x^2} + 4xy+23{y^2})\ .
\end{eqnarray}
 
It is important to note that the on-shell mass corrections 
are proportional to the gauge boson mass squared $m^2$, 
similar to the results obtained in Ref.~\cite{hlz} for 
a scalar mass correction,
and not to $M_S^2$ as found in \cite{wrong}. Consequently, there is no hard
quadratic dependence on the cutoff. On the other hand, 
the proportionality factor $\lambda^{n+2}$ appearing in the 
above expressions implies that the results are rather 
sensitive to the precise value of the cutoff in comparison 
to the string scale $M_S$. This is an intrinsic uncertainty
for any process involving virtual KK graviton exchanges
in an effective theory.

Before we end this section, a few remarks related to the
potential uncertainties of the results are in order:
\begin{enumerate}
\item{Eqs.~(\ref{seagull}), (\ref{rainbow0})
and (\ref{rainbowm}) appear to be infrared divergent 
as a consequence of the pole at $y=0$ for massless gravitons. 
This must be an artifact of 
the fixed-order perturbative calculation 
since gravity is known to be infrared safe.
We will regularize the infrared singularity using the principles 
of dimensional regularization and perform the minimal
subtraction.
The procedure is described in Appendix B.}
\item{It is interesting to 
investigate the limit for a large value of the cutoff 
$\lambda M_S\rightarrow\infty$, or equivalently
$r\rightarrow 0$. As an illustration we consider $n=2$. 
The seagull diagram is independent of $r$ and gives
\begin{equation}
\Pi_S(0)=0,\quad 
\Pi_S(m^2)={\frac{7\lambda^{4} m^2}{18 \pi}}\ ,
\label{sea}
\end{equation}
and the rainbow diagram gives
\begin{equation}
\lim_{r\rightarrow0} \Pi_R(0)=-{\frac{\lambda^{4} m^2}{72 \pi}} ,\quad
\lim_{r\rightarrow0} \Pi_R(m^2)=-
{\frac{29\lambda^{4} m^2}{72 \pi}}\ .
\label{rain}
\end{equation}
Therefore, the total self-energy correction 
$\Pi(p^2)=\Pi_R(p^2)+\Pi_S(p^2)$ takes the following values
at $p^2=0$ and $p^2=m^2$: 
\begin{equation}
\lim_{r\rightarrow0} \Pi(0)=\lim_{r\rightarrow0} \Pi(m^2)= 
-{\frac{\lambda^{4} m^2}{72 \pi}} \ .
\label{tot}
\end{equation}
One would hope that the self-energy corrections vanish as the string 
scale is set to infinity, as required by the decoupling theorem. 
This has been explicitly shown not to be the case by 
the naive results in Eq.~(\ref{tot}).
The problem lies in the fact that an unknown cosmological constant
can also contribute to the gauge boson self-energies via gravitational
interactions. 
A non-zero cosmological constant term is of the form
\begin{equation}
 \Lambda \int d^4 x \sqrt{-g},
\end{equation}
and would lead to an additional contribution to the self-energy
by tadpole diagrams,
\begin{equation}
\Pi_{\Lambda}(p^2) \sim m^2 
{\frac{\Lambda\lambda^{n-2}}{ M_S^{4}}} 
\int_0^1 y^{{\frac{n}{2}-2}} dy \  .
\label{PiL}
\end{equation}
This term can drastically change the previously
calculated self energies. If $\Lambda\sim (\lambda M_S)^4$,
Eq.~(\ref{PiL}) could be at the same order of Eq. (\ref{tot})
and could thus provide the appropriate counter-term.
However, due to the lack of a consistent way of 
determining the cosmological constant, 
the precise value of the term in Eq. (\ref{PiL}) is not known.
As a result, the self-energies and subsequently, corrections
to the $W$ and $Z$ boson masses have inherent uncertainties.
}
\item{In our calculations for the gauge boson self-energies,
we have adopted the momentum-cutoff scheme to regularize the
divergent mass sum and the momentum integral, since we consider
this scheme a most direct reflection of the ultraviolet
behavior. We anticipate that the physics results would not
depend upon the specific regularization scheme.
}
\end{enumerate}

It is important to emphasize that the above potential 
ambiguities are of no concern to us 
if we adopt the $STU$-formalism since the oblique
parameters are manifestly finite \cite{para}.  
The regulator independence built into the definition of the oblique
parameters make the infrared and ultraviolet 
divergences as well as the irrelevant constants drop
 out when taking the difference of the appropriate
combination of self-energies, as we will present next. 
\newpage
\section{Oblique parameters}

The studies of electroweak radiative corrections have
proven to be powerful in constraining new 
physics \cite{para,lynnetal,KL,epsi,MR} 
beyond the SM.
A convenient parameterization of new physics from a higher scale
is the $STU$-formalism \cite{para}. The $S$, $T$ and $U$ parameters
can be obtained by evaluating the self-energy corrections 
at the energy scales $m_Z^{}$ and 0. We write the oblique
parameters as\footnote{These definitions are the same as in 
Ref.~\cite{MR}. In the case under consideration, they are 
identical to those originally introduced in Ref.~\cite{para}.}
\begin{eqnarray}
\alpha\, S &=& 4 s^2 c^2\, {\frac{\Pi_{Z}(m_Z^2)-\Pi_{Z}(0)}{m_Z^2}}\ ,\\ 
\alpha\, T &=& {\frac{\Pi_{W}(0)}{m_W^2}}-{\frac{\Pi_{Z}(0)}{m_Z^2}}\ ,\\ 
\alpha\, (S+U) &=& 4 s^2 \, {\frac{\Pi_{W}(m_W^2)-\Pi_{W}(0)}{m_W^2}}\ ,
\end{eqnarray}
where $s$ and $c$ are the sine and cosine of the weak mixing angle and 
$\alpha$ is the fine structure constant, all measured at $m_Z$. 

The $S$ ($S+U$) parameter measures the difference in the contribution 
of new physics to neutral (charged) current processes
at different energy scales. $U$ is generally small.
The $T$ parameter serves as a comparison between the new 
contributions to the neutral and charged current processes 
at low energy, proportional to $\Delta\rho$.
Comparing experimental data mainly from the LEP and SLC
to SM predictions with $m_H^{}=300$ GeV 
leads to the bounds \cite{constraints},
\begin{eqnarray}
S&=&-0.30\pm0.13\ , \nonumber \\
\label{STU}
T&=&-0.14\pm0.15\ , \\
U&=&\ \ \,0.15\pm0.21\ . \nonumber
\end{eqnarray}

\begin{figure}[t]
\centerline{\psfig{file=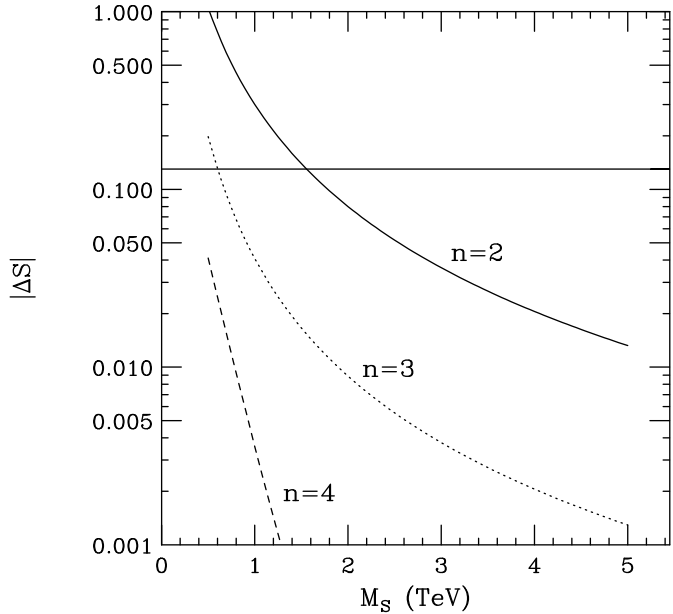,width=3.5in,angle=90}}
\bigskip
\caption[]{ $|\Delta S|$ for $n=2,3$ and 4  
as a function of $M_S$. The region above the line $|\Delta S|$=0.13 
is excluded by experiment at the 1$\sigma$ level.   }
\label{S}
\end{figure}

\begin{figure}[h]
\centerline{\psfig{file=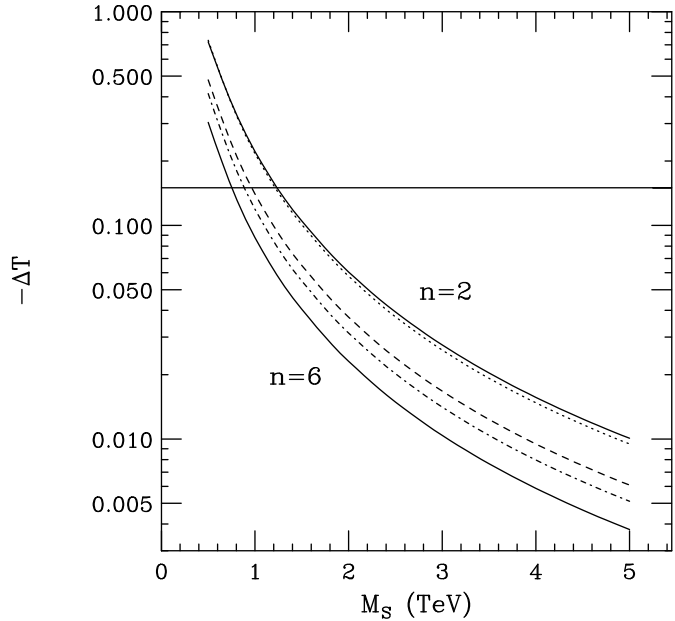,width=3.5in,angle=90}}
\bigskip
\caption[]{ $-\Delta T$ vs $M_S$. The region above the line 
$-\Delta T$=0.15 is excluded at the 1$\sigma$ level. 
Constraints are imposed on $M_S$ for all $n$.}
\label{T}
\end{figure}

\begin{figure}[h]
\centerline{\psfig{file=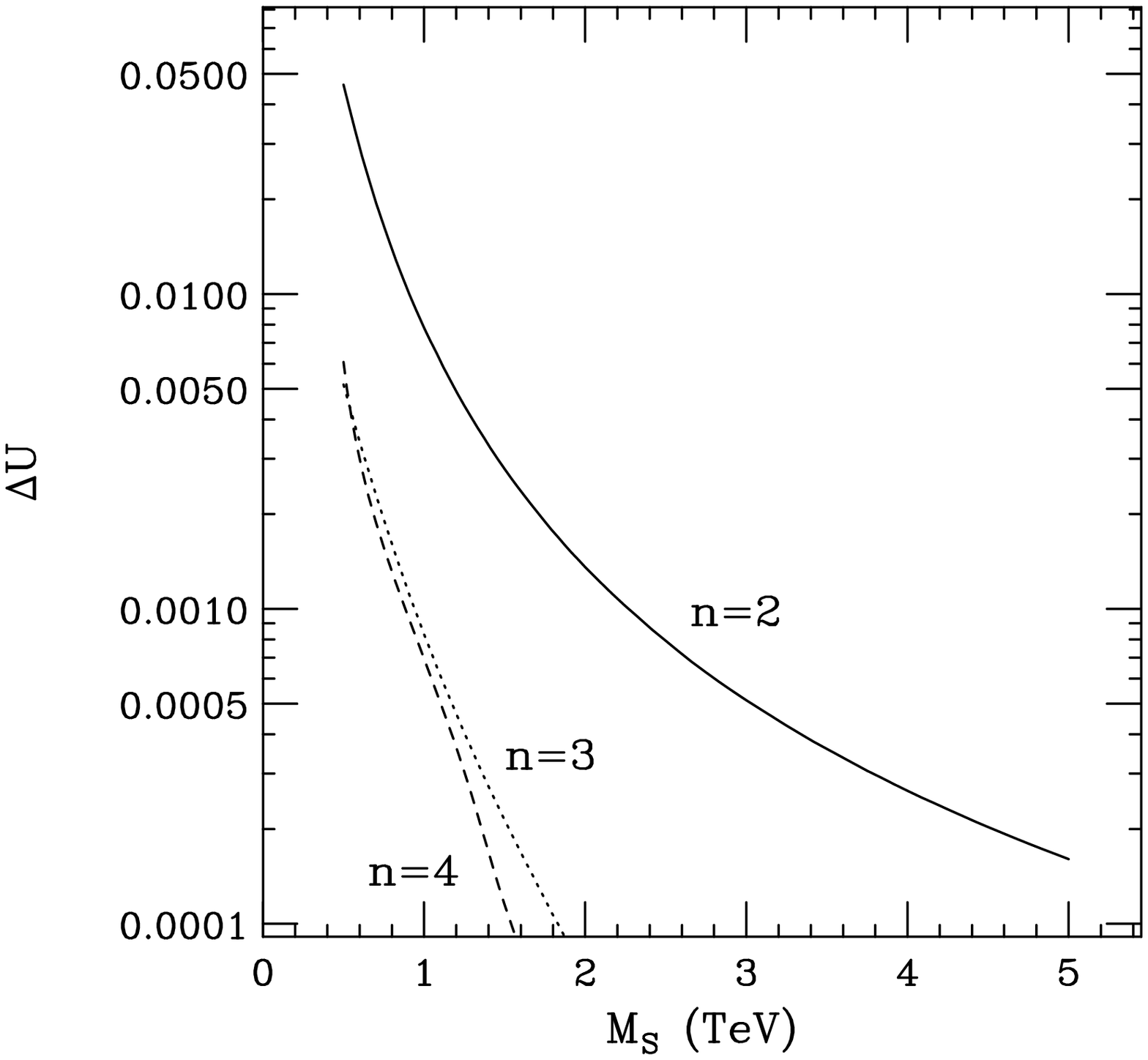,width=3.5in,angle=90}}
\bigskip
\caption[]{ $\Delta U$ vs $M_S$ leads to no constraint.}
\label{U}
\end{figure}

We perform an oblique correction analysis using the calculations 
of the previous section and the SM parameters \cite{constraints}
\begin{eqnarray}
m_Z=91.1867\ \gev\ ,\quad m_W=80.315\ \gev\quad {\rm and}\quad
s^2=0.232\ . \nonumber
\end{eqnarray}
We assume the central values of Eq.~(\ref{STU}) to be the 
SM predictions and attribute the error bars to the physics
contribution of our current interest. Our numerical results 
are presented in Figs.~(\ref{S})-(\ref{U}) where we have 
set $\lambda=1$.
 
In Fig.~\ref{S},
we have plotted $\Delta S$, the excess contribution to the SM 
value of the $S$ parameter arising from the KK graviton exchanges 
of the one-loop self-energy diagrams,
for $n=2$ (positive) and $n=3,4$ (negative).
These values must lie within the 
error bars stated in Eq.~(\ref{STU}) to not be in conflict with precision
electroweak measurements. We see that for $n=2$, this leads to a lower 
bound $M_S>1.55$ TeV and for $n=3$, $M_S>600$ GeV at the $1\sigma$ level. 
There is no constraint for
$n\ge4$ from the $S$ parameter and we have therefore neglected to plot the 
cases $n=5,6$.

Figure~\ref{T} shows that the $T$ parameter imposes constraints on $M_S$ 
for all $n$. This seems to be counter-intuitive at first sight 
since the interactions of KK gravitons with matter should respect 
the custodial $SU(2)$ symmetry and thus lead to a null contribution
to $T$. However, the mass difference of the gauge bosons make their
couplings to the gravitons different, resulting in
a negative graviton contribution to $T$.
In fact, the constraints for $n\ge 3$ are more stringent than that
obtained from the $S$ parameter. For instance,
the lower limit on $M_S$ is raised to 1.25 (0.75) TeV for $n=3$
(6). 

The $U$ parameter is generally small and it is true here
as well. On the other hand, the error bars on $U$ are still
rather large and we would not expect improvement by it.
Indeed, from Fig.~\ref{U}, we see that the $U$ parameter 
places no constraint whatsoever. 
\section{Discussion and Conclusion} 

We have obtained significant constraints on the string scale $M_S$,
 by considering the
$S$, $T$ and $U$ parameters from precision electroweak data. 
One may wonder if the $STU$-formalism is suitable since
gravitons can be lighter than gauge bosons. We
consider our treatment appropriate because as a collective
contribution, the relevant effects are characterized by
the string scale at about a TeV. A more subtle question
is whether the non-oblique corrections would also be as
important and how they can be incorporated. In fact, leading
KK graviton corrections appear even at tree level and 
significant effects were found in the literature \cite{virtual}.
We thus view our approach to single out the oblique corrections
as the next-to-leading contribution as reasonable.

As we see from Eq.~(\ref{tot}), 
the self-energy
corrections remain finite in the limit $M_S\rightarrow\infty$. As explained
in the paragraph following it, it is possible to
achieve ``decoupling behavior'' by introducing a cosmological constant
as a counter-term. 
By no means is this necessary as far as 
the oblique parameters are concerned.
The effect of the heavy states does decouple 
in $S$, $T$ and $U$ as seen from the relation Eq.~(\ref{tot}),
\begin{equation}
\lim_{{M_S}\rightarrow \infty}\, S, T, U = 0\ .
\end{equation}
Altough demonstrated specifically for the case $n=2$, we have confirmed 
this to be true for all $n$. It is reassuring to obtain 
the ``decoupling'' relations, which imply that the radiative
corrections based on this effective field theory of KK gravitons
are under control. 

For all of our numerical analysis, the choice $\lambda=1$ 
corresponds to taking the cutoff scale as the string 
scale. This is the same as the choice in most phenomenological
studies \cite{hlz,virtual}. There is the intrinsic 
uncertainty due to the ratio parameter $\lambda$, although
it is reasonable to assume it to be of order unity.
More definitive results will depend on details of a
string model \cite{strings2} near the string scale.

In summary, we obtained significant lower bounds on $M_S$ 
from the $S$ and $T$ parameters. For one and two standard deviations,
they are
\begin{equation}
\begin{array}{ccccc}
n   &  &  M_S(\gev)\ {\rm at}\ 1\sigma &  & M_S(\gev)\ {\rm at}\ 2\sigma  \\
\hline
2   &  &  1550                 &  & 1100                      \\
3   &  &  1250                 &  &  850  \\
4   &  &   950                 &  &  650     \\
5   &  &   900                 &  &  600     \\
6   &  &   750                 &  &  500     
\end{array}
\end{equation}
These results are comparable to that inferred from current LEP II 
experiments \cite{virtual} and are slightly weaker than 
those anticipated at future runs of LEP and the Tevatron.
\vspace{0.25in}
\acknowledgments 
We thank C.~Goebel, J. Lykken and D. Zeppenfeld for discussions.
This work was supported in part by a DOE grant No. DE-FG02-95ER40896 
and in part by the Wisconsin Alumni Research Foundation.
\vspace{0.25in}
\appendix
\section{}

In this appendix  we summarize the Feynman rules used in the
self-energy calculations. 

The propagator for the massive spin-2 KK 
states $\th^\vn_{\mu\nu}$ is \cite{hlz}
\begin{equation}
i\,\Delta^\th_{\{\mu\nu,\vec{n}\}, \{\rho\sigma,\vec{m}\}}\ (k) 
\ =\ {i\,\delta_{\vec{n},-\vec{m}}\ B_{\mu\nu,\rho\sigma}(k)
\over k^2-m^2_{\vec{n}}+i\varepsilon}\ ,
\end{equation}
where 
\begin{eqnarray}
B_{\mu\nu,\rho\sigma}(k) &=& 
\left(\eta_{\mu\rho}-{k_\mu k_\rho\over m_\vn^2}\right)
\left(\eta_{\nu\sigma}-{k_\nu k_\sigma\over m_\vn^2}\right)
+\left(\eta_{\mu\sigma}-{k_\mu k_\sigma\over m_\vn^2}\right)
\left(\eta_{\nu\rho}-{k_\nu k_\rho\over m_\vn^2}\right)\nonumber\\
&& - {2\over3}\left(\eta_{\mu\nu}-{k_\mu k_\nu\over m_\vn^2}\right)
\left(\eta_{\rho\sigma}-{k_\rho k_\sigma\over m_\vn^2}\right)\ .
\label{B}
\end{eqnarray}

\begin{figure}[tbh]
\centerline{\psfig{file=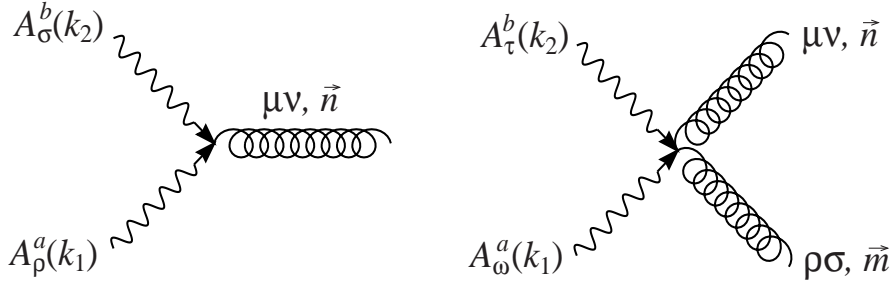,width=12cm,height=4cm}}
\bigskip
\caption[]{The three-point and four-point vertices. 
The KK states are plotted in helices. }
\label{vertices}
\end{figure} 

The three-point vertex is shown in Fig.~\ref{vertices} and the
corresponding Feynman rule is \cite{hlz}
\begin{equation}
i{\frac{\kappa}{2}}\delta^{ab}\left[(m^2+k_1\textbf{.}k_2) 
C_{\mu\nu,\rho\sigma}+D_{\mu\nu,\rho\sigma}(k_1,k_2)\right]\ ,
\end{equation}
where
\begin{equation}
C_{\mu\nu,\rho\sigma}=\eta_{\mu\rho}\eta_{\nu\sigma}+\eta_{\mu\sigma}
\eta_{\nu\rho}-\eta_{\mu\nu}\eta_{\rho\sigma}
\end{equation}
is the tensor that appears in the massless graviton propagator in the de 
Donder gauge, and
\begin{equation}
D_{\mu\nu,\rho\sigma}(k_1, k_2)=
\eta_{\mu\nu} k_{1\sigma}k_{2\rho}
- \biggl[\eta_{\mu\sigma} k_{1\nu} k_{2\rho}
  + \eta_{\mu\rho} k_{1\sigma} k_{2\nu}
  - \eta_{\rho\sigma} k_{1\mu} k_{2\nu}
  + (\mu\leftrightarrow\nu)\biggr]\ .
\end{equation}

The Feynman rule for the four-point vertex as shown in Fig.~\ref{vertices} 
is
\begin{equation}
i{\frac{\kappa^2}{4}}\delta^{ab}\left[(m^2+k_1\textbf{.}k_2) 
C_{\mu\nu,\rho\sigma\mid\omega\tau}+H_{\mu\nu\rho\sigma\omega\tau}(k_1,k_2)+
I_{\mu\nu\rho\sigma\omega\tau}(k_1,k_2)\right]
\end{equation}
where
\begin{equation}
C_{\mu\nu,\rho\sigma\mid\omega\tau}={\frac{1}{2}}\left[\eta_{\mu\omega}
C_{\rho\sigma,\nu\tau}+\eta_{\sigma\omega}C_{\mu\nu,\rho\tau}+
\eta_{\rho\omega}C_{\mu\nu,\sigma\tau}+\eta_{\nu\omega}C_{\mu\tau,\rho\sigma}-
\eta_{\omega\tau}C_{\mu\nu,\rho\sigma}+(\omega \leftrightarrow \tau)\right]\ ,
\end{equation}
\begin{eqnarray}
H_{\mu\nu\rho\sigma\omega\tau}(k_1,k_2)&=&-\biggl[C_{\mu\nu,\rho\tau}
k_{1\sigma}k_{2\omega}
+C_{\mu\nu,\rho\omega}k_{1\tau}k_{2\sigma}-C_{\mu\nu,\omega\tau}k_{1\rho}
k_{2\sigma}+(\rho \leftrightarrow \sigma)\biggr]\nonumber\\
&&-\biggl[(\mu,\nu) \leftrightarrow (\rho,\sigma)\biggr]\ ,
\end{eqnarray}
and
\begin{eqnarray}
&&~~I_{\mu\nu\rho\sigma\omega\tau}(k_1,k_2)=\\
&&\Biggl\{\biggl[(C_{\sigma\omega,\nu\tau}-\eta_{\sigma\tau}\eta_{\nu\omega})
k_{1\mu}k_{2\rho}+(C_{\nu\omega,\sigma\tau}-\eta_{\sigma\omega}
\eta_{\nu\tau})k_{1\rho}k_{2\mu}+(\mu \leftrightarrow \nu)\biggr]+
\left(\rho \leftrightarrow \sigma\right)\Biggr\} 
+C_{\mu\nu,\rho\sigma}k_{1\tau}k_{2\omega}\nonumber\ .
\end{eqnarray}
\newpage
\section{}

We provide details of the treatment of the infrared
singularity for the case $n=2$. From the expression for the self-energy
contribution from the seagull diagram, Eq.~(\ref{seagull}), 
we extract the integral
\begin{eqnarray}
\int_0^1\, {\frac{x\,{y^{-3 + {\frac{n}{2}}}}}{x+y}}\,
 \left( x + 3\, y \right)\left( x + 6\, y \right)\, dy & = &
 \left({\frac{2\,x^2}{n-4}}\right) 
{}_2F_1\left[1,\,(-4 + n)/2,\,(-2 + n)/2,\,-1/x\right]\,
 \nonumber \\ & &
 + 18\,\left( {\frac{x}{n-2}} +\, 
     {\frac{1}{n}}{}_2F_1\left[1,\,n/2,\,
         (2+n)/2,\,-1/x\right]
      \right)
\end{eqnarray}
where
\begin{equation}
_2F_1\left[a,b,c,D\right]=\sum_{k=0}^{\infty} {\frac{(a)_k (b)_k}{(c)_k}}
{\frac{D^k}{k!}} 
\end{equation}
is the hypergeometric function and $(a)_k$ is the Pochhammer symbol
\begin{equation} 
(a)_k=\Gamma(a+k)/\Gamma(a)\ . 
\end{equation}
This result is convergent only for $n>4$. For  $n=2+\epsilon$, 
we analytically continue the above equation 
and expand it in a power series in $\epsilon$, 
\begin{equation}
{\frac{16x}{\epsilon}}-x\biggl[x-\ln(1+x^{-1})\biggr]+{\cal O}(\epsilon)\ .
\end{equation}
Integrating the ${\cal O}(\epsilon^0)$ term of the above
expression over $x$, we obtain
\begin{equation}
-\int_0^1 dx~
x\biggl[x-\ln(1+x^{-1})\biggr]\ =\ {\frac{14}{3}}\ .
\end{equation}
We can repeat the procedure for $\Pi_R(0)$.
The ${\cal O}(\epsilon^0)$ term of the integral in  
Eq.~(\ref{rainbow0}) is
%
%
%
\begin{eqnarray}
&&30{r^2}\biggl[Li_2({\frac{1}{1 - r}}) - 
     Li_2({\frac{2}{1-r}}) + 
     Li_2(-r^{-1}) \biggr]
+3{r^2}\biggl[r\ln {\frac{r}{r+1}} - 
     10\ln{\frac{2(r+1)}{r-1}} \biggr] \nonumber \\ 
&&\qquad +{3\over2}r  \left( 2r-1+ 40\ln 2 \right)  -{\frac{1}{3}}\ ,
\end{eqnarray}
%
where $Li_2$ 
is the dilogarithm function. The same procedure 
can be carried out for the  integral in  $\Pi_R(m^2)$, but 
the expressions are exceedingly cumbersome and not at all 
illuminating. We do not show them here. 

The point we wish to emphasize is that by performing 
dimensional regularization in the number of extra spacetime dimensions $n$, we
are able to systematically isolate the infrared divergence and calculate 
finite quantities with a sensible physical meaning. Using the results in this
Appendix, we showed that the oblique corrections vanish 
identically in the limit $M_S\rightarrow\infty$, as one would expect. 

We have
demonstrated this procedure only for $n=2$, but it can be carried out for 
any $n$. For example, for $n=3$, we would substitute $n=\epsilon+3$ and so on.

\end{document}